# Towards the bio-personalization of music recommendation systems: a single-sensor EEG biomarker of subjective music preference

**Dimitrios A. Adamos*[1,3], Stavros I. Dimitriadis[2,3] and Nikolaos A. Laskaris[2,3]**

*Asterisk indicates corresponding author*

[1]School of Music Studies, Aristotle University of Thessaloniki, 54124 Thessaloniki, Greece
[2]AIIA Lab, Department of Informatics, Aristotle University of Thessaloniki, 54124 Thessaloniki, Greece
[3]Neuroinformatics GRoup, Aristotle University of Thessaloniki, Greece - http://neuroinformatics.gr

Correspondence to:

Dr. Dimitrios A. Adamos
School of Music Studies, Faculty of Fine Arts,
Aristotle University of Thessaloniki,
GR-54124 Thessaloniki, GREECE
Tel: +30 2310 991839
Fax: +30 2310 991815
E-mail: dadam@mus.auth.gr

## Abstract

Recent advances in biosensors technology and mobile electroencephalographic (EEG) interfaces have opened new application fields for cognitive monitoring. A computable biomarker for the assessment of spontaneous aesthetic brain responses during music listening is introduced here. It derives from well-established measures of cross-frequency coupling (CFC) and quantifies the music-induced alterations in the dynamic relationships between brain rhythms. During a stage of exploratory analysis, and using the signals from a suitably designed experiment, we established the biomarker, which acts on brain activations recorded over the left prefrontal cortex and focuses on the functional coupling between high-β and low-γ oscillations. Based on data from an additional experimental paradigm, we validated the introduced biomarker and showed its relevance for expressing the subjective aesthetic appreciation of a piece of music. Our approach resulted in an affordable tool that can promote human-machine interaction and, by serving as a personalized music annotation strategy, can be potentially integrated into modern flexible music recommendation systems.

# 1 Introduction

Novel forms of communication between human brain and computing machines are anticipated as brain activity decoding pertains a wide range of applications that extend far beyond restoration in physically impaired people. High expectations about the benefits of this symbiosis have already been set by the cross-fertilization of relevant ideas across theorists, scientists and innovators [43]. Meanwhile, the detection of subject's intentions and mental states are additionally motivated by the newly born market of brain-oriented consumer applications [4]. Moreover, the recording of brain activity within naturalistic environments is gradually entering into the current neuroimaging picture [31].

Among contemporary brain imaging methodologies, the electroencephalography (EEG) appears as a convenient non-invasive technique that records the electrical activity of the brain using multiple electrodes placed on the scalp. EEG measures voltage fluctuations resulting from the summation of synchronous extracellular currents; these are due to ionic flows generated by large populations of neurons that share a similar spatial organization. The spectral content of EEG traces is of critical importance. It reflects the ongoing activity resulting from the locally synchronized neuronal activations. Common themes in EEG studies are the nonstationarity of the brain activity and the existence of inherent oscillations usually referred to as *brain rhythms.* Brain rhythms are identified by means of band-pass filtering within specific frequency bands (an example is depicted in Figure 1). These rhythms reflect oscillatory activity that spreads among distinct neural structures and shape functional connectivity and cognitive processing [9].

Recent advances in signal analysis accredit EEG as an invaluable neuroimaging tool with high temporal resolution [34] and the incorporation of low-cost wireless EEG sensors in multimodal mobile cognitive monitoring tasks has been initiated (see [11] and [49]). In this context, several possibilities have been identified for exploiting someone's affective states, as these are decoded by means of real-time processing of EEG signals [31]. Among them, the convenient bio-assessment of music-liking level emerges as a technological achievement that would majorly enhance current music recommendation systems.

Functional neuroimaging studies have already shown that music can effectively modulate subcortical structures of the human brain and strongly affects the human disposition and emotional state [28]. Changes in regional cerebral blood flow were identified during an exceptionally strong level of positive emotional engagement to music listening [7], namely "chills", "thrills" or "shivers down the spine" [23]. At the level of cortical activations, affective responses during music listening have been associated with asymmetrical frontal EEG activity (see [3], [44] and [45]). In the majority of studies in this domain (e.g. [3], [32] and [45]), the assessment of music affectiveness has been influenced by the emotional polarization (positive/negative) induced by the two-dimensional [37] valence-arousal model. However, sad-sounding music should not be treated as undesirable as it occasionally can be the subject's own choice [23]. Hence, as a remedy to the previous approach, the model of subjective liking/disliking has been incorporated in latest EEG studies (as in [44], [20], [21] and [29]. Additionally, almost all studies ([3], [20], [21], [23], [32] and [44]) involved "active" music listening in their experimental procedures; the participants had to decide on their affective ratings during EEG recordings and indicate their scores upon listening to the corresponding musical excerpts.

Lately, studies of functional connectivity have consistently identified 'network hubs', i.e. brain regions which are critically important for efficient neuronal communication (for a review see [22]). Their functional role has been recognized in a wide range of cognitive tasks, and it is known to manifest itself in the form of dynamical coupling within and between functional networks. To address the functional conjunction among different brain rhythms, researchers have recently started to examine the phenomenon of cross-frequency coupling (CFC). CFC is considered among the principal mechanisms that facilitate the interactions between local and global networks, integrate distributed information and subserve the cognitive faculties (see [25] and [26]). The functional role of transient cross-frequency coupling among brain rhythms has been demonstrated in many recent studies. When triggered (or perturbed) by external sensory input or internal cognitive events, it may lead to multimodal dynamic patterns with particular measurable properties at the level of recorded signal [10]. CFC is evident in at least two forms: (1) phase synchrony, during which a consistent number of high-frequency cycles occur within single cycles of a lower frequency rhythm [46] and (2) phase-amplitude coupling (PAC), during which the phase of a low-frequency rhythm modulates the amplitude of a higher-frequency oscillation [47]. Both CFC modes can facilitate neural communication and neural plasticity [18]. As a more technical introduction to the concept of PAC, we have included the example of a synthetic signal in Figure 2. The PAC attributes of this signal incorporate a phase-locking behavior between the phase of a 2 Hz sinusoid and the amplitude of a higher frequency sinusoid (10 Hz).

Inspired by the established significance of CFC, we hypothesized that musical aesthetic appreciation may be mediated and conveyed by means of neural representations associated with similar coordination phenomena as well. This working hypothesis was particularly motivated via a recent work, in which increased functional connectivity of the frontal cortex with subcortical dopaminergic areas was reported [53] in the case of highly rewarding music. Based on this hypothesis, we envisioned the integration of a metric of this appreciation into modern flexible music recommendation systems as a means of enhancing user's feedback and rating processes with bio-personalization features.

This study was designed so as to detect the putative CFC correlates of subjective music appreciation and incorporate them in a reliable, computable biomarker. The followings summarize the main characteristics of our approach.

a) Two passive listening paradigms (i.e. there is no task for the subject, as in [7] and [8]) were employed. The recordings from first/second paradigm were used for establishing/validating the biomarker.
b) By focusing on the assessment of spontaneous aesthetic brain responses, the implications of active cognitive processing during music listening were circumvented.
c) The limitations of emotional polarization (positive/negative) posed by the valence-model related strategies (as in the assessment of sad-sounding music [23]) were avoided.
d) Our investigations were confined to CFC-characteristics within the same recording site, so as to lead to a metric that would depend upon brain activity recorded from a single sensor. This satisfied the need for computational efficiency and compatibility with low-cost consumer devices.
e) The suggested biomarker operates in a personalized manner tailored to the user. Hence, it is not confronted with *inter-subject variability* (i.e. the well-known fact that across subjects there are considerable variations in the anatomical and functional organization of the cortex ([1],[2],[19])).

Our results show that cross-frequency-coupling (CFC) measurements can be effectively formulated as a single-sensor personalized biomarker that would facilitate the personalized tagging of subjectively attractive music as a way of cognitive monitoring in a real-life situation. With convenience as main characteristic, it is potentially useful in naturalistic HCI schemes suitable for user-friendly mobile and Internet applications.

The remainder of this paper is structured as follows. Section 2 serves as an introduction to the notion of phase-amplitude CFC and describes the methodological route for establishing the proposed Biomarker. Section 3 describes briefly the adopted experimental paradigms. Section 4 is devoted to the presentation of the obtained results and is followed by the discussion and concluding section.

## 2  Methods

*2.1  Modulation Index (MI): a phase-amplitude cross-frequency coupling measure*

In the present study, we aimed at exploring the changes in PAC-level among brain rhythms recorded at a single-sensor and induced by listening to music. To quantify PAC we employed the modulation index (MI), a new measure for the empirical assessment of phase-amplitude functional coupling [47]. In the followings we first introduce the algorithmic steps involved in the computation of MI index using the synthetic signal of Figure 2 and then provide a more formal definition of this index.

Figure 3 outlines the steps for quantifying the MI index for a particular pair of frequencies. The user defines the frequency range of these components and the algorithm starts by extracting them using band-pass filtering. Then the trace of instantaneous phase of the low frequency component (Fig. 3b) and the trace of instantaneous envelope (Fig. 3c) of the high-frequency component are derived. Next, the phase range [-π, π] is partitioned into a user-defined number of intervals (or 'bins') and the instantaneous phase values of the low-frequency component (Fig. 3d) are grouped accordingly (i.e. "binning" ). The temporally-associated envelope samples of high-frequency component are then grouped based on the membership of the corresponding phase values (Fig. 3e). Finally, by within-bin averaging of the envelope samples, and a suitable normalization, probabilities are assigned to each bin (Fig. 3f). In this way an empirical distribution is formed, the deviation of which from the uniform distribution serves as a measure of the coupling between the low and high frequency components. Overall, the algorithm outputs a probabilistic index (ranging between 0 and 1) that indicates the strength of the detected PAC, with higher MI values indicating stronger coupling.

Stating the previous in a more formal manner, the MI estimation begins by defining two frequency bands of interest, referred to as the "phase-modulating" and the "amplitude-modulated" signals. We denote these two frequency bands, based on their central frequencies, as $f_1$ and $f_2$ for the modulating signal $x_1(t)$ and the modulated signal $x_2(t)$ respectively. Next, we describe the algorithmic steps required for the computation of MI index from the original (i.e. unfiltered) sensor signal $x(t)$. Figure 4 accompanies this description by exemplifying the steps using a signal from our recordings.

**step_i:** The band-limited signals $x_1(t)$ and driven $x_2(t)$ are formed.

**step_ii:** Using Hilbert transform, the analytic signals $z_1(t) = a_1(t)e^{j\theta_1(t)}$ and $z_2(t) = a_2(t)e^{j\theta_2(t)}$ are derived.

**step_iii:** By pairing the instantaneous phases from the former signal with the amplitude envelope from the latter, the bivariate time series $\Theta A(t) = [\theta_1(t), \alpha_2(t)]$ is constructed containing the amplitudes of the $f_2$ brain rhythm at each phase of the modulating rhythm.

**step_iv:** After selecting a reasonable number of bins $N_{bins}$, the set of instantaneous phases $\theta_1(t)$ is partitioned into equally-sized intervals. The binning of the first component of $\Theta A(t)$ defines the binning of the second (i.e. the samples of $\alpha_2(t)$ are grouped according to their associated time indices).

**step_v:** By within-bin averaging, amplitude-related values are computed $\{A_2(j)\}_{j=1:N_{bins}}$ for the second component of $\Theta A(t)$.

**step_vi:** These values are transformed to discrete probabilities $\{P_2(j)\}_{j=1:N_{bins}}$ by normalizing with the overall sum over the bins

$$P_2(j) = \frac{A_2(j)}{\sum_{k=1}^{N_{bins}} A_2(k)} \quad (1)$$

It is important to notice that these $\alpha_2(t)$-related quantities are conditioned by the $\theta_1$-dependent binning. If there is no phase-amplitude coupling between the involved frequencies ($f_1, f_2$), the distribution $\mathbf{P_2}=\{P_2(j)\}$ is uniform, hence the amplitude envelope of $x_2(t)$ has no dependency on the instantaneous-phase signal $\theta_1(t)$. Hence, the existence of phase-amplitude coupling can be signaled by a deviation from the uniform distribution. To this end, the Kullback-Leibler (KL) divergence is adopted. KL is a standard statistical dissimilarity measure for comparing two distinct distributions; in our case $\mathbf{P_2}$ against the corresponding uniform distribution $\mathbf{U}$. The formula reads as follows:

$$D_{KL}(\mathbf{P_2}, \mathbf{U}) = H(\mathbf{U}) - H(\mathbf{P_2}) = \log(N_{bins}) - H(\mathbf{P_2}) \quad (2)$$

$$\text{where } H(\mathbf{P_2}) = -\sum_{j=1}^{N_{bins}} P_2(j) \log(P_2(j)) \quad (3)$$

$H$ stands for the Shannon entropy of a distribution, which takes the maximal value log($N_{bins}$) in the case of uniform distribution. The definition of MI-index incorporates the following normalization (since $H(\mathbf{P_2}) \leq \log(N_{bins})$)

$$MI = MI(x_1(t), x_2(t)) = \frac{D_{KL}(\mathbf{P}, \mathbf{U})}{\log(N_{bins})} \quad (4)$$

As a consequence, the MI index ranges between 0 and 1. Greater deviations of the $\mathbf{P_2}$ distribution from the uniform distribution (as inferred by the KL distance) lead to higher MI values, while an MI value near zero indicates the absence of phase-amplitude coupling [47].

For the purpose of this study, the sensor signal was narrow-band filtered within multiple bands of 1-Hz width, having their central frequencies spaced regularly over the [1-40] Hz range. Then the MI-value was repeatedly estimated for all possible couplings using $N_{bins}$ =18. The particular selection was suggested in [47] and corresponds to dividing the 360° interval into sectors of 20°.

## 2.2 CFC-based indices of subjective music preference

Considering that music is a mild stimulus and that it may result only to feebly alterations in the brain waves recorded over the scalp, we decided to examine more delicate descriptors that go beyond spectral characterization and can encapsulate dynamic behavior across neural networks in the human brain [10]. We formed the hypothesis that aesthetic responsiveness to music would be reflected in the way the inherent oscillatory components are coupled within each other when listening to the music. To verify this, based on experimental data, we formed comodulograms. These are maps of MI-based estimates and their construction includes multiple scanning across a range of frequencies for possible couplings. Working for each recording site independently, we computed, via eq(4), the MI-value for all possible pairs $(f_1, f_2) = (f_i, f_j)$, $f_i, f_j \in \{1, 2, ... 40\}$ Hz. The derived measurements were tabulated in a [40×40] matrix $^{sensor}\mathbf{MI}$ with entries $M(i,j)$ corresponding to a particular pair of frequencies $(f_i, f_j)$ and quantifying the amplitude modulation of $f_j$-related rhythm by the phase of $f_i$-related rhythm.

As our aim was to detect music-induced alterations, we selected to work with contrasting indices of the form

$$NMI(i,j) = \frac{MI_{music}(i,j) - MI_{rest}(i,j)}{MI_{rest}(i,j)} \quad (5)$$

Hence, in each entry of a $^{sensor}\mathbf{NMI}_{[music]}$ matrix, the relative change of CFC strength was stored with respect to a baseline condition (resting state). We refer to such matrices as music-related comodulograms and use the subscript to denote the piece of music that gave rise to them. In addition, the associate superscript indicates the recording site.

*2.3 Contrasting music related comodulograms*

In an exploratory stage of this work, we systematically searched for the particular rhythms (i.e. frequency pairs) between which the changes in the functional coupling best reflect the subject's music preference. To this end, we derived music-related comodulograms stemmed from listening to a "neutral" piece of music and compared them with comodulograms emerged during listening to the favorite piece of music. The tabulated CFC scores were denoted as $^{sensor}\mathbf{NMI}_{[neutral]}$ and $^{sensor}\mathbf{NMI}_{[favorite]}$ respectively. Having available these CFC patterns for N subjects, namely $\{\mathbf{NMI^k}\}_{[neutral]}$ and $\{\mathbf{NMI^k}\}_{[favorite]}$, k=1,2,..N we adopted a discriminant-analysis perspective [51]. We considered each (i,j) entry in the comodulogram patterns as a feature and measured its descriptive power for separating the neutral-music related CFC patterns from the ones induced by the preferred music. We employed the ratio of between-scatter and within-scatter as a feature-ranking criterion and estimate the following score.

$$J(i,j) = J(f_i, f_j) = \frac{\left|\overline{\mathrm{NMI}}_{[favorite]}(i,j) - \overline{\mathrm{NMI}}_{[neutral]}(i,j)\right|^2}{\mathrm{Var}_{[favorite]}(i,j) + \mathrm{Var}_{[neutral]}(i,j)} \qquad (6)$$

$$\text{where } \overline{\mathbf{NMI}}_{[music]} = \frac{1}{N}\sum_{k=1}^{N}\mathbf{NMI}^k_{[music]},$$

$$\mathrm{Var}_{[music]}(i,j) = \frac{1}{N-1}\sum_{k=1}^{N}\left|\overline{\mathrm{NMI}}_{[music]}(i,j) - \mathrm{NMI}^k_{[music]}(i,j)\right|^2$$

In this way, a [40×40] matrix $^{sensor}J$ was derived for each sensor. Within each entry, there was tabulated a score reflecting the discriminability of a particular CFC-pair. These scores were sharing a common numerical scale (the higher the J-value the more distinguishable the music induced patterns) and hence facilitated across sensor comparisons.

## 2.4 Examining temporal concistency

In an attempt to specify a CFC-related biomarker that would not only be sensitive to music preference but also perform steadily, we further extended the previously mentioned computation of separability J by introducing the temporal duration as a varying parameter.

Specifically, we derived a sequence of J maps $\{J_k(i,j)\}_{k=1:5}$, by including signal segments from k=5 time windows $W_k=[0-T_k]$, $T_k=20,40,...100$ sec. Put it into words, the first time window corresponded to 20 sec from the start of music, the second included 20 additional seconds and so on. Using the 5 corresponding J-values associated with a coupling of a particular frequency pair, we estimated for each entry (i,j) the inverse of coefficient of variation (ICV) as follows:

$$ICV(i,j) = ICV(f_i, f_j) = \frac{\bar{J}_k(i,j)}{std(J_k(i,j))} \cdot H(|\bar{J}_k(i,j)| - 0.3) \quad (7)$$

where H(x) denotes the Heaviside step function and was incorporated so as to mask[1] the entries of minute but constant class-separability. The ICV index served as an additional scoring function that helped in the refinement of our class-separability measurements and contributed to the definition of our biomarker.

## 2.5 The CFC Biomarker

The above-mentioned explorations, which were based on Group analysis, led us to define the following biomarker as a score reflecting music liking

$$BM_{[music]} = {}^{AF3}NMI_{[music]}(f_{sel_1}, f_{sel_2}),$$
$$f_{sel_1} = [24-28]Hz, f_{sel_2} = [32-36]Hz \quad (8)$$

The above formula implies that the signal from AF3 sensor (see Fig.5) during listening to a particular piece of music is filtered around 26 Hz and 34 Hz (with bandwidth 4 Hz) and the outputs are used for deriving normalized estimates of PAC (via eq.5).

## 3 Experimental data

Before providing a short description of our experimental design, some notes are in order. First, among the principal objectives of this work was the establishment of a biomarker that would reflect the listener's fondness for a piece of music or song, regardless the emotions induced by it. We were searching for a signature of music preference that would emerge naturally, without the subject being engaged in any cognitive task (like "rate this song"). Second, the recording of brain activity should be performed under naturalistic conditions and accompanied by the minimum possible discomfort. For these reasons we employed a wireless state-of-the-art consumer EEG device, and asked the participants to experience the music just as if it was being broadcasted by a radio station, i.e. without producing any kind of active response. Their opinion about each piece of music had been registered during an independent listening session before the recording.

---

[1] The selection of '0.3' as the masking threshold was dictated by the need to reduce clutter in Fig.7.

## 3.1 Participants

The subjects were recruited from the School of Informatics of Aristotle University of Thessaloniki. They were all students who considered listening to music as among their personal interests and had a very clear positive attitude towards a particular music genre (either rock or pop). They signed an informed consent after the experimental procedures had been explained to them.

## 3.2 Data acquisition

Data were acquired with the *Emotiv EPOC* headset (http://emotiv.com). The EEG recording included 14 active electrodes referenced to the left/right mastoid, with a topological arrangement that can be seen in Fig.5. The signals were digitized at the sampling frequency of 128 Hz, with an effective bandwidth of [0.5 - 45] Hz.

All the recordings and music listening session were carried out in a professional studio environment. *Genelec* (Genelec, Finland - http://genelec.com) active studio monitor equipment was used for delivering audio signals. *Audacity* (http://audacity.sourceforge.net) open source software was used for editing the music and *OpenSesame* [33] open-source software was used for automating the experiment setup. The across-platform (OS X, Unix, Windows) implementation of the synchronization procedure (event-triggering/marking) between *OpenSesame* and Emotiv's EEG recording software suite (Testbench) is described in detail and available online[2] on our group's web site.

---

[2] http://neuroinformatics.gr/research/tutorials

## 3.3 Experiment-A

The first of the two distinct experimental procedures resulted in the necessary data for establishing our biomarker. Fourteen (14) subjects (two females), with an average age of 25 years, participated in a passive music listening task. Each subject, having had selected beforehand his own favorite piece of music, participated in a single music listening task of three parts: a) 2 minutes of silence, b) 2 minutes of listening to a "neutral" song, c) 2 minutes of listening to his favorite song.

The aim of this experiment was to compare the brain activity patterns elicited while listening to intensely pleasing music with the patterns elicited by other music. Using a piece of music that was the subject's own choice was considered the most reliable way to invoke intense positive aesthetic responses [7]. By the same token, each subject identified some pieces of music as *neutral* from a pool of tracks including pieces of genre other than his preference. The scope for such selection was to record brain activity when listening to a piece of music that would invoke a minimal aesthetic response, i.e. would be neither pleasing nor annoying [39].

As a way of establishing a common reference point for all listeners, we selected among the tracks characterized as neutral the one in common for all subjects. That was *Enya's "Watermark" (1998)*, which was delivered to all participants together with their own favorite track. The list of favorite tracks included songs from various artists (like *Siouxsie and the Banshees, Jenifer Lopez*, *Queen* and *Lynyrd Skynyrd* ).

Apart from brain signals during listening to **"favorite"** and **"neutral"** music, spontaneous brain activity was also recorded in the beginning of the listening task. The latter served as an active baseline condition, which reflected brain's **resting state** and could be exploited so as to express music-induced changes in a universal way.

Before placing the headset and starting the recording, the subject sat comfortably in an armchair and the loudness of the speakers was adjusted to a reasonable level. In a single recording session, and after an initial recording of resting state lasting for 2 minutes, the two different music tracks (the favorite and neutral one) were delivered in randomized order with an interleaved period of silence. The subject had been instructed to keep his eyes open and gaze at a cross in front of him. During debriefing at the end of the recording session, each participant confirmed that he had experienced the two types of music without any distraction.

*3.4 Experiment-B*

The second experimental procedure provided additional data for the further justification of the proposed biomarker. Three (3) additional subjects (all males) participated in a second music listening experiment. This experiment was inspired by a recently introduced procedure that was designed having in mind the transactions in contemporary music-recommendation industry [40]. Likewise, we exploited an online music recommendation service and asked the participants to rate a predefined list of several songs using a [0-5] ranking scheme, which was delimited as follows. 0: 'Not my style', 1: 'Feels ok for background music', 3: 'I would like to listen again to this song', 5:'Bookmark this song in my favorite list'. Ranks 2 and 4 were reserved for in-between ratings. After each participant had accomplished his own rating, we selected 12 songs equally distributed among ranks 1, 3 and 5. On a separate recording day, each participant was submitted to a passive listening task involving the random presentation of extracts from the 12 songs. An initial recording of resting state lasting for 2 minutes preceded the recording during listening to the music compilation. The extracts had been defined as audio thumbnails based on the verse-chorus summarization model for rock/pop music [5], following a *section-transition* strategy [14], and assembled into a single audio stream with silent periods of 5 sec in between.

*3.5 Preprocessing*

The preprocessing of multichannel signals included artifact reduction based on independent component analysis (ICA) ([16], [35]). For each continuously recorded dataset (containing EEG during rest and while listening to music), we used EEGLAB [15] to zero the components that were associated with artifactual activity from eyes, muscle and cardiac interference. The estimated mixing matrix was used to reconstruct the multichannel signal from the rest of ICs. The components related to eye movement were identified based on their scalp topography and their temporal course. The components reflecting cardiac activity were recognized from the regularity of rhythmic pattern in time domain and the widespread topography. Muscle activity related ICs were identified based on statistical terms (if the kurtosis of corresponding time courses was higher than a predefined threshold, $kurt_{thr}=12$), spectral characteristics and topographies (if temporal brain areas were included) [17]. Hence, in the following results, the use of "ICA-cleaned" brain activity is implied.

# 4   Results

Due to the inherent nature of our work (i.e. exploratory data analysis), extensive experimentations were carried out based on the introduced methodological framework. We scrutinize multiple dimensions like recording condition, electrode location and elapsed time. In addition, we performed statistical testing and control based on surrogate data and a randomization process. We also performed a comparison with the standard approach of using spectral characteristics, based on estimates of power spectral density of EEG traces. To ease the presentation of the obtained results, we have decided to adopt the indices of section 2 (i.e. J-values and ICV) and collectively summarize the bulk of information in the form of graphs where the consistent trends are easily spotted. More specifically: a) Figure 5 depicts the significant changes in PAC among brain rhythms due to music listening. b) Figure 6 includes separability maps that reflect, at group level, the differences between the comodulograms of favorite and neutral music. c) Figure 7 includes the ICV-maps reflecting the consistency of separability in time. In addition, the results from the statistical tests and the separability scores corresponding to spectral characteristics are provided as supplementary material.

## 4.1   Defining the Biomarker

First, using the signals from experiment-A, we derived comodulograms from different recording conditions (rest and listening) and started comparing them so as to identify the sensors and the frequency pairs that clearly reflect the influence of music as alterations in the coupling between brain rhythms. Figure 5 exemplifies this step, by illustrating the MI measurements derived from signals recorded at two different sensors. The first column corresponds to resting condition while the middle column to passive listening. The point-by-point differences between the two maps are shown in the rightmost column and clearly indicate that the music exerts a diffused pattern of influence that includes both increases and decreases in the functional coupling. Each row corresponds to a different sensor and it can be seen that CFC-estimates may differ among sensors (the results corresponding to the statistical testing of these CFC-estimates are depicted in supplementary figure S1). The observation of such comodulograms made evident that any possible systematic behavior would have been obscured by measurement imperfections and the ever-present inter-subject variability.

To alleviate this, we decided to form a normalized MI measurement (eq. 5) so as to express music-induced alterations in a sensor-adaptive and subject-referenced manner. Moreover, we resorted to a pattern recognition strategy for feature ranking (see methods section) and estimated the descriptive power of each NMI measurement regarding the problem of classifying "neutral"/"favorite" music from brain-activity signals. This stage of analysis resulted in a set of "universal" maps; they epitomized, at a group-analysis level, all individual comparisons and portrayed the discriminability of each sensor and frequency-pair. Figure 6 includes the set of "neutral vs favorite music"-separability maps that was derived from brain activity during music listening for 80 sec. The corresponding statistical significance of the discriminability had been assessed by means of a randomization procedure as described in the corresponding section of supplementary material (see Fig.S2). It becomes evident that the NMI values from various frequency pairs and different sensor locations can be considered as useful features for discriminating neutral from favorite music. For instance, the increase of NMI values at $^{AF3}$(24Hz, 34Hz) and the corresponding decrease at $^{P7}$(12Hz, 37Hz) are associated with a high separability level (0.52) that signifies, respectively, the trend of increased and decreased functional coupling during listening to favorite music. For comparison purposes, we also estimated the corresponding class-separability level of power spectra density (PSD) estimates from the same signals (the results, expressed as J scores, are shown in the supplementary figure S3). Interestingly, the highest discriminability levels were identified at particular sensors/frequencies (e.g. $^{AF3}$(4Hz), $^{FC5}$(5Hz) and $^{F4}$(23Hz)), in which PSD was higher during the favorite music. However, the maximal level of J index was 0.34, well below the level achieved via CFC measurements.

A definite selection was not feasible based on the results shown in Figure 6, since there were many spots of similar intensity that appear of equivalent importance to the classification task of music preference. We then decided to bring the temporal dimension under consideration and examined the separability from the perspective of variations in time (see Section 2.4, eq.7). Figure 7 contains the corresponding ICV maps for all sensors. By taking into consideration the stability over time, a much more consistent picture emerged that led us to the suggestion of biomarker in eq.8.

Figure 8, contains the biomarker values for all individuals and various time windows. It reflects the most prominent changes induced by music, as a function of the time elapsed from the beginning of music. It can be observed that after the 40 first seconds the BM-index

starts, systematically, to score higher the brain activity during favorite music. In particular, after being exposed for 80 sec to music, 13 out of 14 subjects were characterized by BM-values that were higher for favorite music (i.e. values in red bars are higher than corresponding blue ones, for T=80).

*4.2 Biomarker validation*

Using the cortical signals from experiment-*B* we computed, independently for each one of the three participants, the BM values for all the extracts in the defined music compilation. As an index of congruence between the subject's ratings (about each of the 12 included tracks) and the derived BM-scores, we estimated the Spearman's rank correlation coefficient ($\rho$) as a function of the time elapsed from the beginning of each track. Figure 9 includes the aggregated results (mean across subjects) and also provides an indication about the inter-subject variability (through standard deviation). A systematic behavior (i.e. low std) of the coefficient $\rho$, across subjects, is observed when incorporating more than 40 seconds of music-modulated cortical activity. A correlation between subjective evaluation and the introduced biomarker exists from the beginning and achieves its highest value at the latency of 70-seconds, with the mean $\rho$ exceeding 0.8.

# 5  Discussion and conclusion

Brain processes during music listening do not simply involve a cognitive consolidation of isolated musical entities (i.e. sounds), but rather an intangible blending of dynamical brain states induced by temporal-based computations of musical patterns (see [30], [50] and [27]). To cope with this complexity, we exploited cross-frequency coupling between brain rhythms as a means of capturing the information exchange via synchronization / desynchronization phenomena. Inspired by recent arguments about the crucial role of CFC in the organization of large-scale networks (as in [10] and [26]), we designed a biomarker that would be derived from the interplay of oscillatory cortical rhythms recorded at a single site and reflect subjective music preference. Our index builds over the observation that the phase of high-$\beta$ band oscillations influences the amplitude of low-$\gamma$ band oscillations in a reproducible way that is detectable over the left prefrontal cortex (AF3 sensor).

It has already been reported in [7], [12], [40], [48] and [52] that the neural underpinnings of aesthetic music experience include the participation of the frontal lobes, and in particular the prefrontal cortex. Furthermore, dopamine has been identified as a rewarding factor encoding this appreciation during music listening in [38] and [40]. In this respect, it has been shown that interconnected frontal cortical and subcortical dopaminergic areas increase their functional connectivity as a form of musical reward and are highly significant during musical processing in [53]. Recent fMRI studies ([24] and [40]) have supported the relation of prefrontal areas with music, memory and emotion, while the dorsal medial prefrontal cortex was revealed as the brain hub that links the three of them [24]. Additionally, medial frontal activations while listening to music have been linked to the perceptual context in [36] and emotion perception in [32] and [41]. The engagement of the left frontal cortex area is also in agreement with previous works (see [3], [42] and [44]) reporting EEG asymmetries regarding the pre-frontal activations due to aesthetically pleasing (positive) stimuli.

Among the related literature there are only two recent articles ([20] and [21]) that deal with the problem of decoding music preference from EEG signals and meanwhile share the vision for implementation under real-life situations. However, their work differs from ours in many aspects. Regarding the experimental design, active listening was employed (participants had to respond 'like'/ 'dislike' during listening). Regarding signal analysis, characteristics based on the modulation of brain rhythms were extracted across the scalp and fed to a common -across subjects- classifier. Despite the methodological differences, their empirical results pointed to the role of $\beta$ and $\gamma$ bands as well.

In our study, the adopted experimental paradigms were designed so as to register -on a personalized level- the subjective music preferences during passive music listening, having in mind an online streaming music service that would introduce songs to the user and implement assistive decisions accordingly (e.g. skip a song of minimum subject-aesthetic interest / tag a song in my 'watch list' etc.). Respectively, the suggested biomarker can be incorporated as a user's feedback and rating process within the framework of contemporary music recommendation systems [13]. Conventionally such a feedback is collected either explicitly or implicitly. In the first case, the system acquires the user's preferences by involving him in an interactive unequivocal rating process. In the second case, the feedback is acquired by means of passive monitoring and interpreting user actions (e.g. skipping the current song). The presented biomarker stands somewhere in the middle. It may be incorporated in a user-centric system that continuously collects subjective preferences without interruptions. Alternatively, after establishing a suitable normalization in the range of its values, the biomarker may be exploited in specific modes of existing music recommendation schemes such as user profiling and evaluation. It can be conveniently integrated within content-based filtering methods so as to enhance the user-preference profile with bio-personalization attributes (and extend the repertoire of correlations with audio content-based features). It can also be exploited by a user-centric evaluation process to assess the 'success' of a recommendation (e.g. song or music-genre).

In the upcoming human-centered digital age, the novel process of *prosumption* (resulting from the blending of *producer* and *consumer* actions) is anticipated to reform human-computer interactions [6]. Online music listening practices of users will be fed back to recommendation services and are expected to drastically shape the selection of music that the listeners would eventually experience. This opens a new application field for low-cost neuroimaging devices to implement the 'bio-monitoring' of the listener's aesthetic experience and introduce bio-personalization properties in future music recommendation services. The introduced biomarker, with the necessary adaptations towards real-time scenarios, fits ideally in this picture.

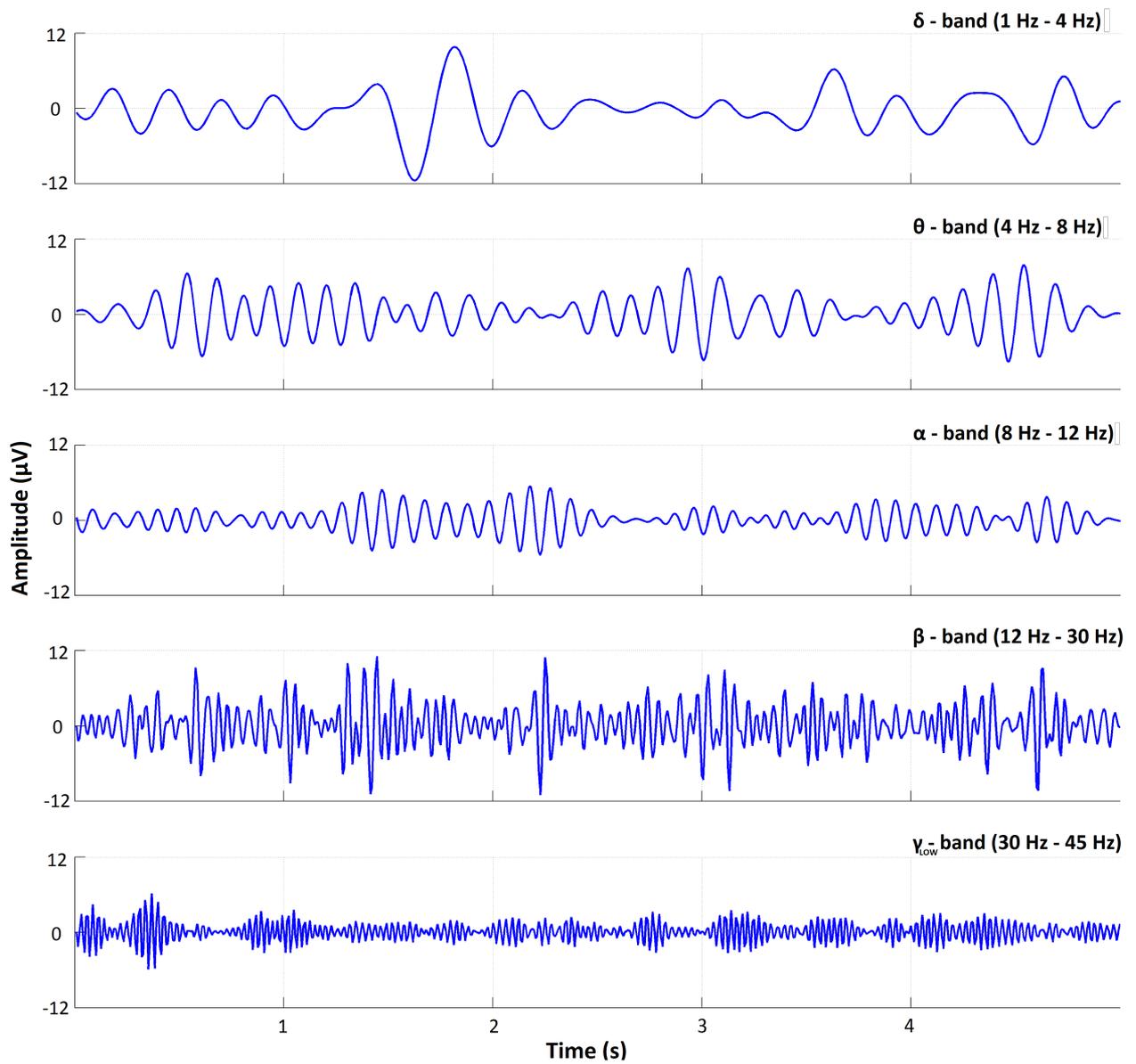

Figure 1. An example of EEG signal decomposition to its constituent brain rhythms (δ, θ, α, β, low-γ). The shown traces correspond to spontaneous brain activity recorded at a site (electrode F7) over the left frontal cortex.

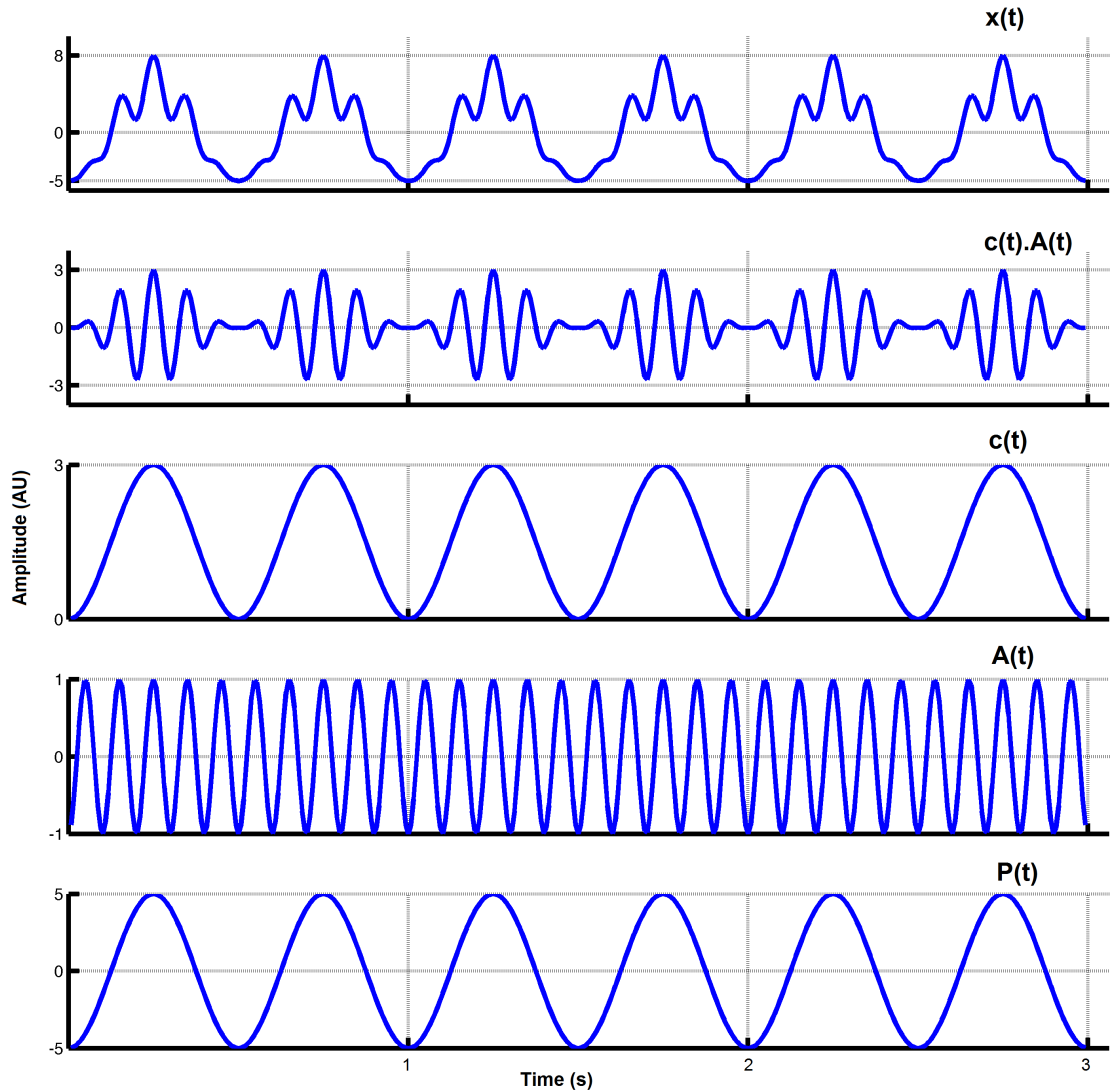

$$x(t) = c(t) \cdot A(t) + P(t) = 3\,\frac{\sin(4\pi t) + 1}{2} \cdot \sin(20\pi t) \;+\; 5\sin(4\pi t)$$

Figure 2. A composite signal exemplifying the PAC concept. It has been built from a phase-modulating rhythm (2 Hz) and an amplitude-modulated rhythm (10 Hz). The intermediate steps for forming x(t) from the oscillatory components (P(t), A(t) and c(t)) are depicted in distinct rows. The steps have been adapted from [47]

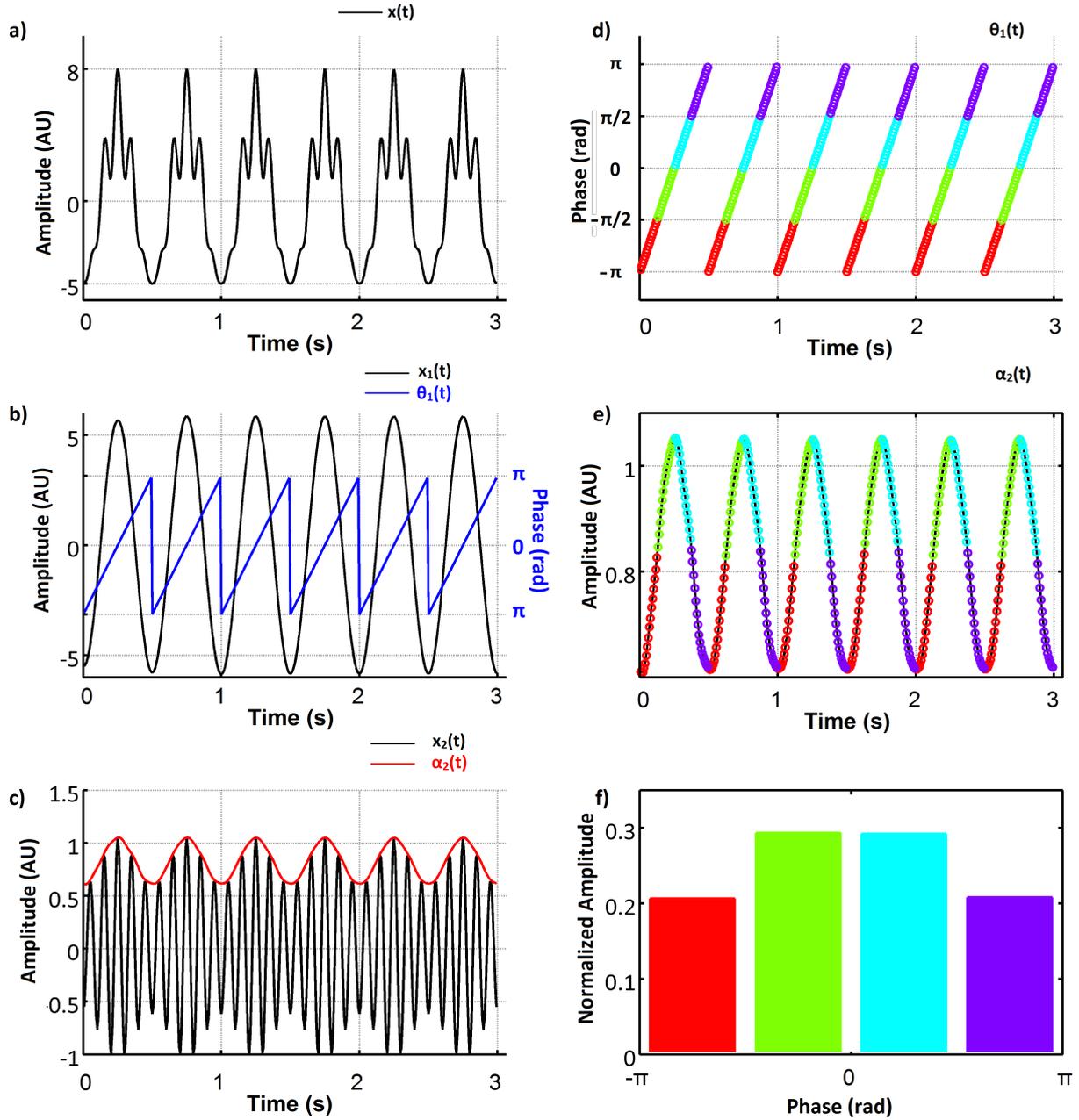

Figure 3. Untangling the phase-amplitude coupling of the synthetic signal depicted in Figure 2 by means of Modulation Index: a) the signal x(t) b) the low frequency component of the signal, band-pass filtered around 2 Hz and its instantaneous phase c) the high-frequency component of the signal, band-pass filtered around 10 Hz and its amplitude envelope d) the radian phase range of [-π,π] is partitioned into N=4 bins and the instantaneous phase values of the low frequency component are binned respectively as indicated by the color-code e) the corresponding (temporally associated) samples of the envelope of the high-frequency component are similarly binned f) By within-bin averaging of the previous envelope values and normalization, probabilities are assigned to each bin. The deviation of this distribution from the uniform distribution indicates the strength of the underlying phase-amplitude coupling.

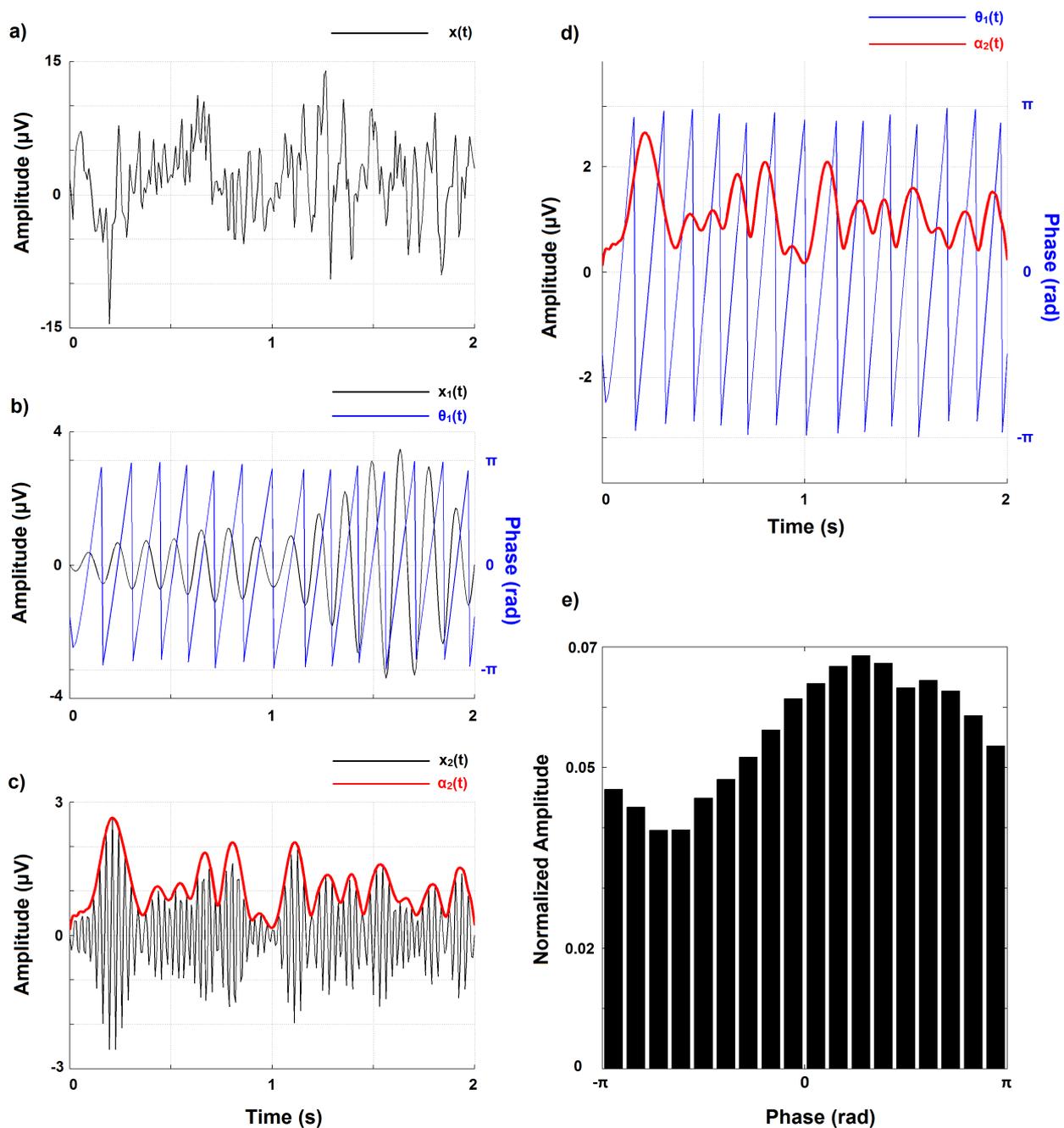

Figure 4. Demonstrating the steps for estimating the phase-amplitude cross-frequency coupling using an EEG trace from AF3-electrode: a) The raw (1-40 Hz) signal x(t) b) The low-frequency (6-8Hz) signal $x_1(t)$ and its instantaneous phase $\theta_1(t)$, c) The high-frequency (30-34Hz) signal $x_2(t)$ and its amplitude envelope $\alpha_2(t)$, d) The joint-analysis of $\theta_1(t)$ and $\alpha_2(t)$ (i.e. the binning of the former is used to group the amplitude values of the latter) results to the histogram shown in e). The estimated MI-level is 0.0053.

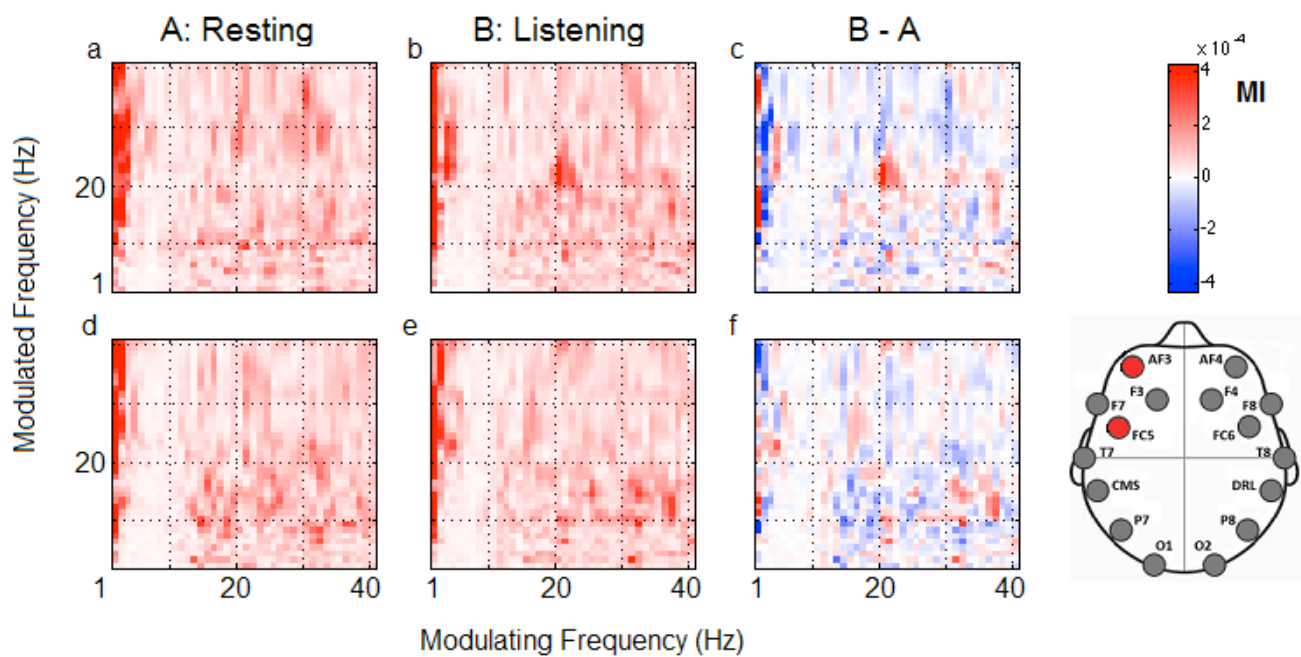

Figure 5. Changes in cross-frequency-coupling during passive listening to music. Comodulograms from 60 seconds of brain activity recorded at electrodes AF3 (top row) and FC5 (bottom row) from subject 1, during resting state (a,d) and music listening (b,e). Rightmost MI-maps reflect the music induced changes and have been derived via subtraction of the corresponding maps (middle – leftmost ones). A common color code has been used for all panels.

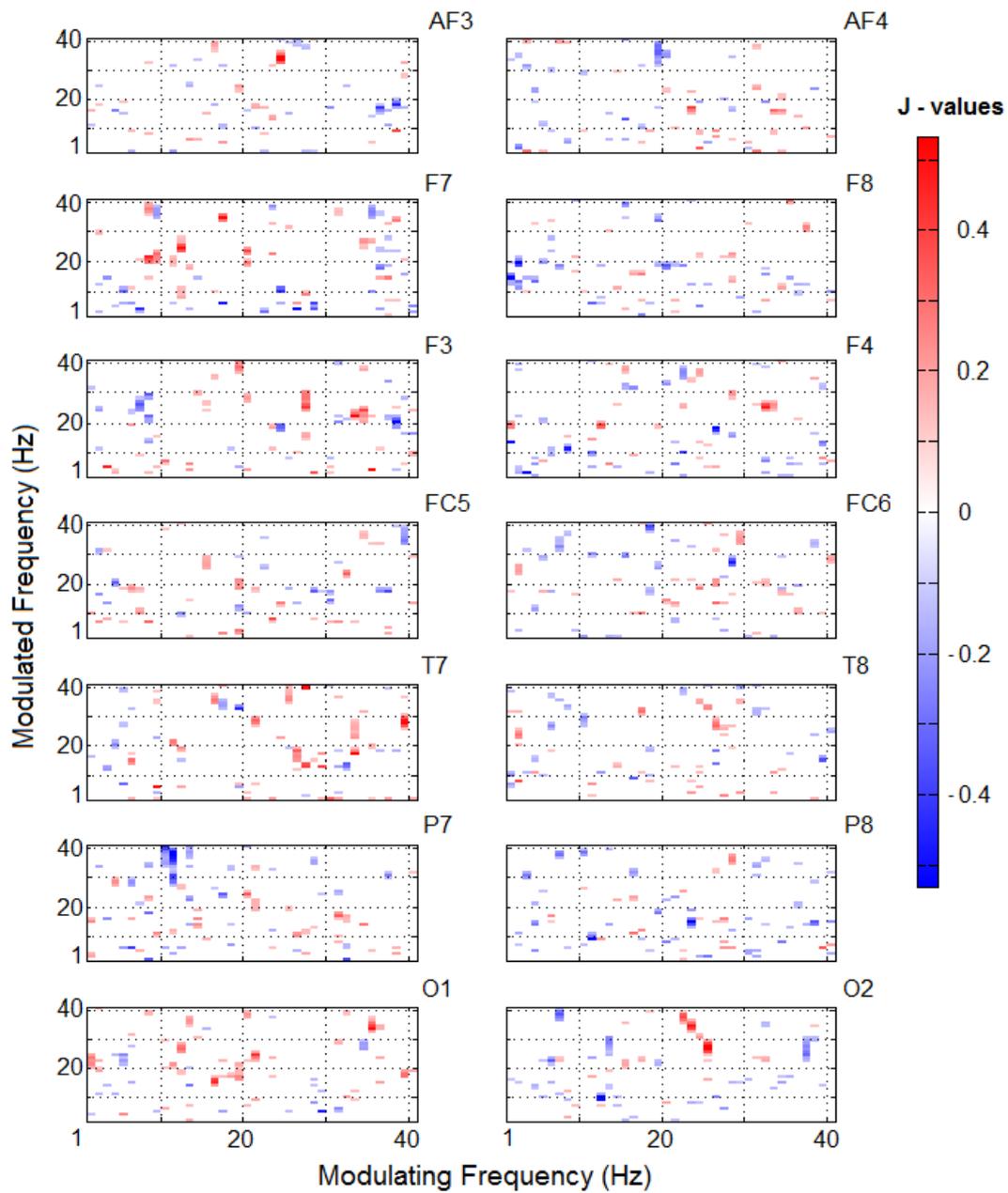

Figure 6. Separability maps depicted over the cross-frequency domain. Each map corresponds to a particular electrode and reflects the differences between neutral and favorite music comodulograms of the first 80 seconds of corresponding recorded brain activity. In addition to separability strength (for clarity, values below 0.1 were thresholded), a bipolar color code has been adopted for denoting the sign of 'favorite'-'neutral' difference. Pixels with bright red(blue) color, indicate frequency-pairs of higher(lower) PAC measure during listening to favorite music.

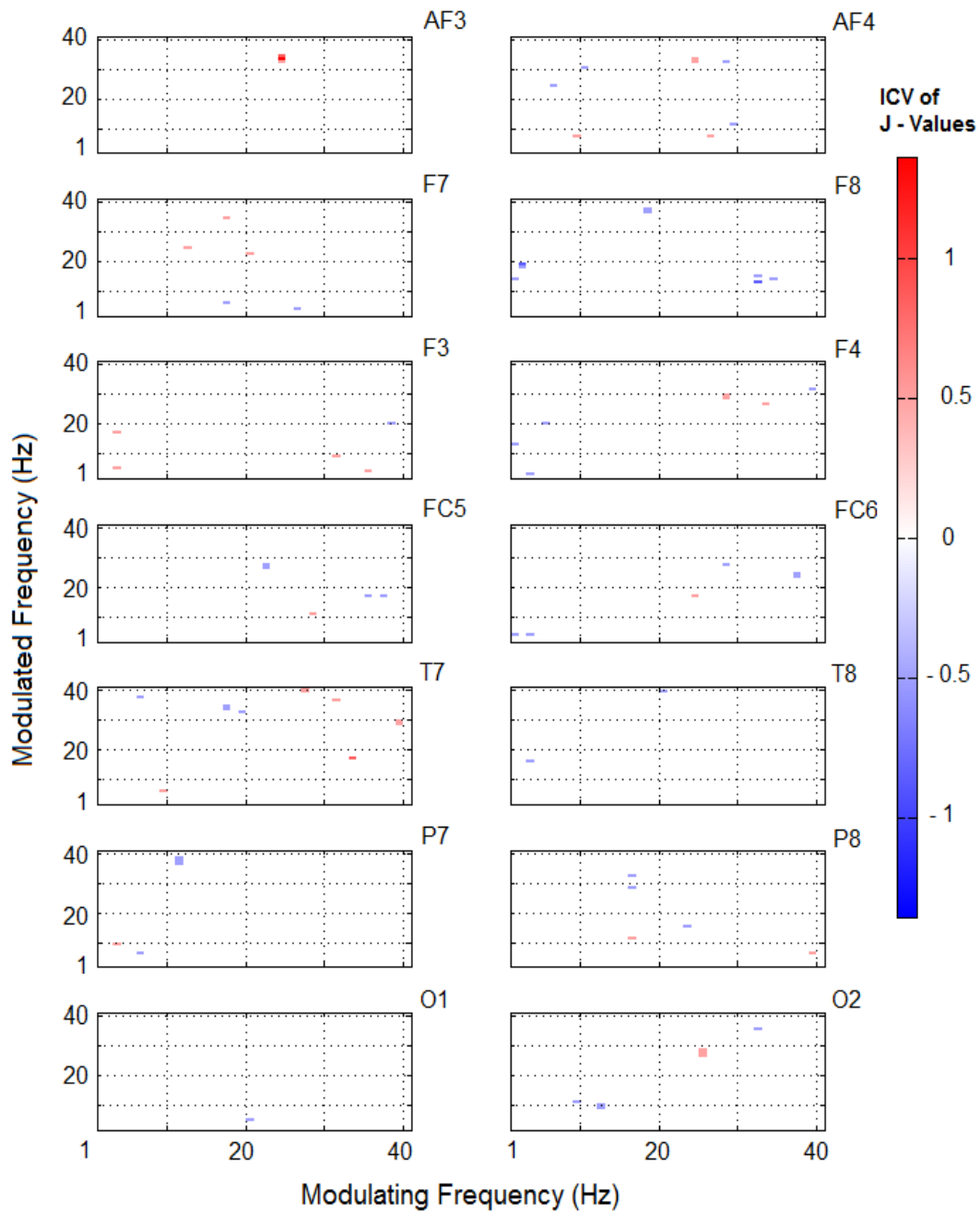

Figure 7. ICV-maps over the cross-frequency domain. Each pixel indicates the consistency over time of the corresponding class-separability index. ICV is estimated over the first 40, 60, 80 and 100 seconds of recorded brain activity. The bipolar color code is borrowed from Fig.6 and denotes the sign of 'favorite'-'neutral' difference in CFC-strength.

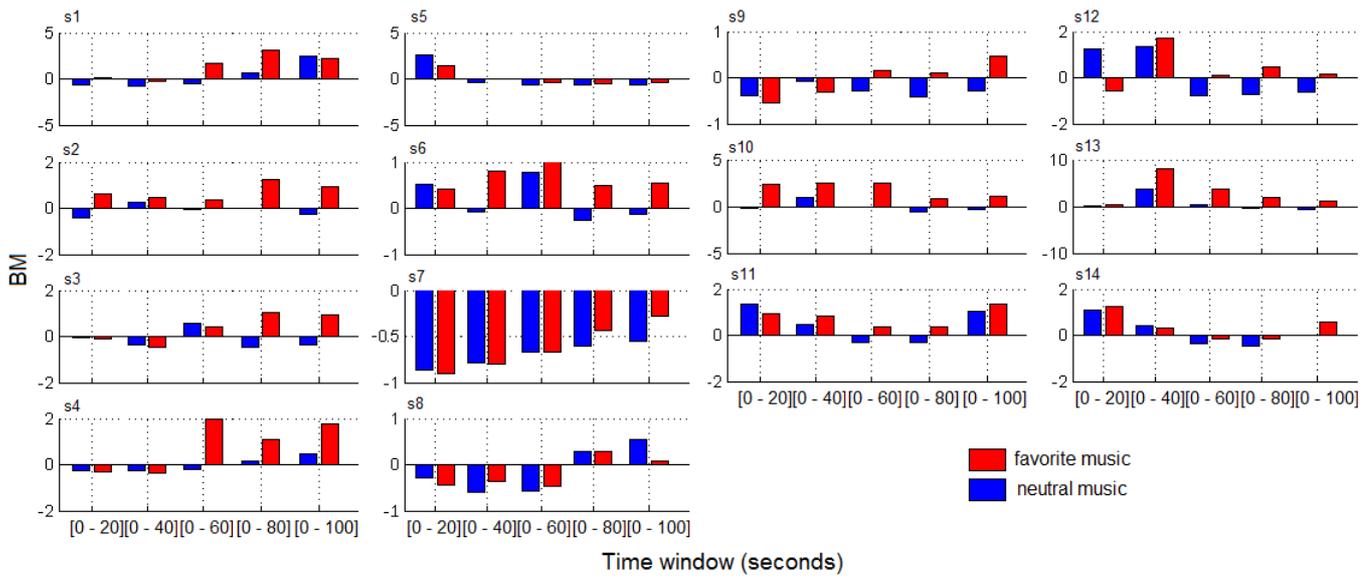

Figure 8. The values of MI-related biomarker for different time windows.

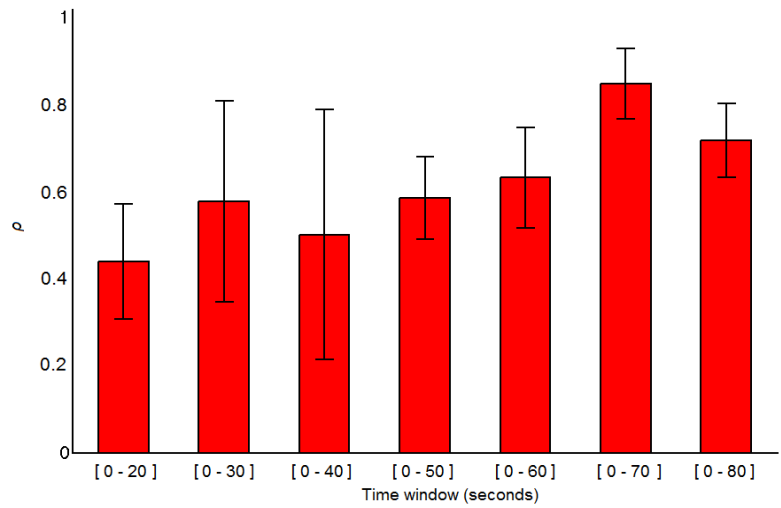

Figure 9. Spearman's rank correlation between BM values and subjective evaluation scores for the music extracts in experiment-*B*.